\documentclass[aps,prx,reprint,longbibliography,nofootinbib,superscriptaddress,floatfix]{revtex4-2}

\usepackage{slashed}
\usepackage{verbatim}
\usepackage[T1]{fontenc}
\usepackage{mathbbol}
\usepackage[dvipsnames]{xcolor}
\usepackage{orcidlink}

\usepackage[normalem]{ulem}
\usepackage[english]{babel}
\usepackage{lipsum}
\usepackage{physics}
\usepackage{dcolumn}
\usepackage{tensor}
\usepackage{comment}
\usepackage{placeins}
\usepackage{graphicx,color,overpic,mathtools}
\usepackage{amsthm,amsmath,amssymb,mathrsfs}
\usepackage{braket,bm,bbm,setspace}
\usepackage{booktabs}
\usepackage{cancel}
\usepackage{float}
\usepackage{xargs}

\definecolor{myred}{RGB}{179, 27, 27}

\usepackage{hyperref}
\hypersetup{
    colorlinks=true,
    linkcolor=myred,
    citecolor=myred,
    urlcolor=myred
}
 
\def\be{\begin{equation}}
\def\ee{\end{equation}}
\def\bea{\begin{eqnarray}}
\def\eea{\end{eqnarray}}
\def\beq{\begin{eqnarray}}
\def\eeq{\end{eqnarray}}

\begin{document}

\title{Nonlinear Dynamics in General Relativity}

\author{Vitor Cardoso\orcidlink{0000-0003-0553-0433}}
\affiliation{Center of Gravity, Niels Bohr Institute, Blegdamsvej 17, 2100 Copenhagen, Denmark}

\author{Jaime Redondo-Yuste\orcidlink{0000-0003-3697-0319}}
\affiliation{Center of Gravity, Niels Bohr Institute, Blegdamsvej 17, 2100 Copenhagen, Denmark}

\author{Ulrich Sperhake\orcidlink{0000-0002-3134-7088}}
\affiliation{DAMTP, Centre for Mathematical Sciences, University of Cambridge, Wilberforce Road, Cambridge CB3 0WA, UK}

\author{Furkan Tuncer\orcidlink{0009-0009-2826-815X}}
\affiliation{Center of Gravity, Niels Bohr Institute, Blegdamsvej 17, 2100 Copenhagen, Denmark}
\affiliation{Bilkent University, Dept. of Physics, 06800 Bilkent, Ankara, Turkey}

\begin{abstract}
Black holes and gravitational waves are consequences of the nonlinear character of the Einstein equations. Yet, the remarkable properties of General Relativity points to the existence of other effects. Here we uncover new nonlinear facets of gravity. We establish higher harmonic generation, spectral broadening and focusing in the Einstein Klein-Gordon system. In vacuum, we show that scattering of monochromatic waves at quadratic order is weakly sensitive to frequency, at large wavelengths. These aspects can both explain the seemingly smooth behavior of mergers, but also caution us against too simplistic an interpretation of waveforms.
\end{abstract}

\maketitle

\noindent \textbf{\textit{Introduction.}} The world is an intrinsically nonlinear system. The sound of snapping fingers, the breaking of ocean waves or water droplet formation, all are manifestations of nonlinear behavior.
Some of the most remarkable consequences of General Relativity (GR) -- such as black-hole (BH) formation and cosmological solutions -- arise from its intrinsically nonlinear character. However, dynamical processes involving BHs {\it seem} to be exquisitely quiet. The well-posedness of the initial value problem assures us that drastic consequences are absent for mild initial conditions~\cite{Foures-Bruhat:1952grw,Ringstrom_2009}, but it does not guide us in the search for interesting nonlinear behavior.
With the exception of BH formation itself, gravitational interactions do not seem to ``agitate'' spacetime too much. For example, the merger of two equal mass BHs gives rise to a smooth merger waveform at large distances, with no sign of significant migration of energy to smaller or larger scales than that set by the binary itself.
Likewise, BH perturbation theory results -- where spacetime is approximately described by the nonlinear geometry of a single BH -- agrees surprisingly well with fully nonlinear results upon the extrapolation to equal-mass objects~\cite{Davis:1971gg,Price:1994pm,Berti:2010ce,Sperhake:2011ik,vandeMeent:2020xgc}. Why is GR so seemingly quiet? 

There are several possible answers to the above concerns. The first is that BHs truly represent the end of short-distance physics, absorbing all small scale (linear and nonlinear) phenomena, which hence remain invisible to external observers. This however, would require a mechanism whereupon no high frequency fluctuation is allowed to travel away from the vicinity of a BH. Another possibility is that GR is indeed turbulent, but BH dynamics is governed by an inverse turbulent cascade, preventing the nonlinear excitation of smaller scales than those driven by e.g. the binary dynamics~\cite{VanRaamsdonk:2008fp, Adams:2013vsa, Green:2013zba, Yang:2014tla, Iuliano:2024ogr, Ianniccari:2025nkf, Kehagias:2025zws, Ma:2025rnv}. This, however, does not fully explain the unreasonable effectiveness of perturbation theory to describe BH mergers. Finally, it is possible that turbulent-like phenomena exist, but remain confined to strong field regions and hence unseen by far-away observers. After all, turbulent phenomena in fluid dynamics are observed in the electromagnetic spectrum, an essentially linear channel that couples to the fluid in the highly turbulent regions. 

Our goal here is to launch a systematic exploration of nonlinear effects in the interaction of pulses in GR. Recently, the phenomenon of frequency-doubling (second-harmonic generation) has gathered interest in the context of BH physics~\cite{Berti:2025hly}. Such second-harmonics (dubbed quadratic quasinormal modes) are present in the higher harmonics of the gravitational-wave (GW) signal emitted when two BHs merge~\cite{London:2014cma, Cheung:2022rbm, Mitman:2022qdl}. The possibility to model the excitation of these higher harmonics~\cite{Redondo-Yuste:2023seq, Perrone:2023jzq, Zhu:2024rej, Khera:2024bjs, Ma:2024qcv, Bucciotti:2024jrv, Bucciotti:2024zyp, Bucciotti:2025rxa, Bourg:2024jme, Bourg:2025lpd, Fransen:2025cgv}, and to detect them with future GW detectors such as LISA~\cite{Yi:2024elj, Lagos:2024ekd}, has triggered a flurry of activity~\cite{Berti:2025hly}. Nevertheless, frequency-doubling in the ringdown is not the only way in which nonlinearities manifest themselves in GR. We begin by systematically studying nonlinear features in the scattering of pulses in the Einstein-Klein Gordon theory in spherical symmetry. Building upon that, we characterise the leading nonlinear behavior in the scattering of GWs off a Schwarzschild BH. Finally, we draw some insights on nonlinearities in BH mergers. 

\noindent \textbf{\textit{Scattering of scalar waves in flat space.}} Consider first the simplest possible setup: a spherically symmetric, massless, real scalar field $\phi$, and use perturbation theory to study the leading nonlinear correction. In spherical symmetry, the metric takes the form~\cite{Bizon:2008iz},
\be
ds^{2}=e^{2\alpha(t,r)}(-e^{2\beta(t,r)}dt^{2}+dr^{2})+r^{2}d\Omega^2\,,
\label{eq:metric_ansatz}
\ee
where $d\Omega^2$ is the round metric on the 2-dimensional unit sphere. We also define the mass function (following conventions of~\cite{Bizon:2008iz}) as $m(t,r)=(1-e^{-2\alpha})r$, and assume small initial data $\phi(0,r)=\varepsilon u_0(r)$, and $\partial_t\phi(0,r)=\varepsilon u_1(r)$, with $\varepsilon\ll 1$. 
\begin{figure*}[t!]
\includegraphics[width=0.32\linewidth]{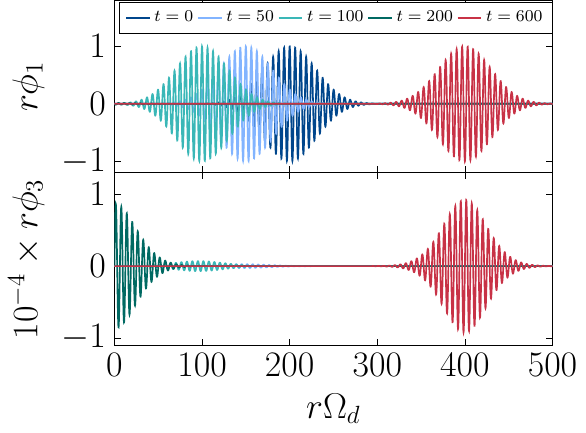}
\includegraphics[width=0.32\linewidth]{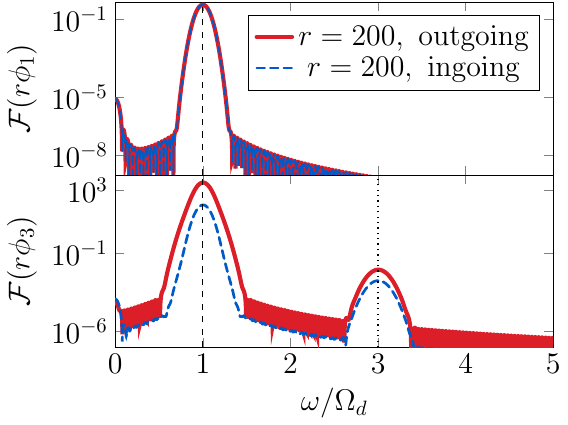}
\includegraphics[width=0.33\linewidth]{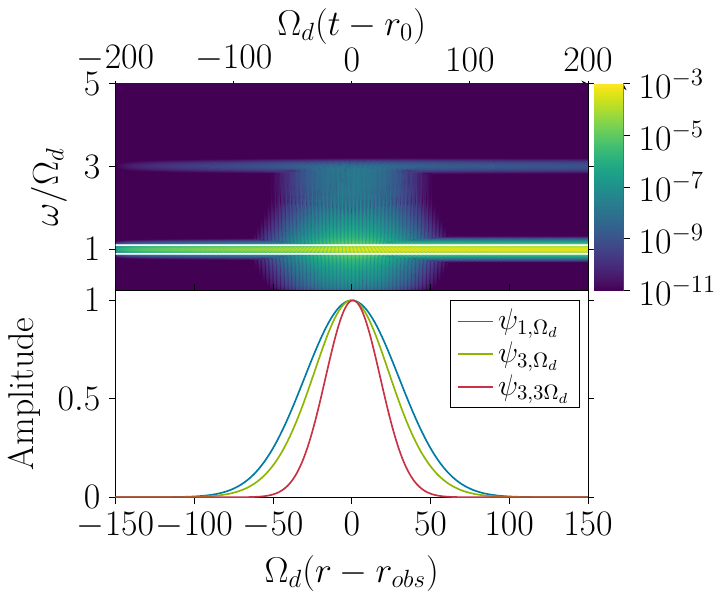}
  \caption{\textbf{Left:} Profile of the linear $r\phi_1$ and first nonlinear correction $r\phi_3$ at different times (see colors), for quasi-monochromatic initial data $\Omega_1=\Omega_2=\Omega_3=\Omega_d=1$ with $r_0=200,\sigma=30$. The bulk of the nonlinear field is generated in the focusing region $r\sim 0$, and spatial focusing is apparent. \textbf{Center:} Fourier transform of the field of the left panel observed at $r=200$ at early times (blue), and late times (red), demonstrating the excitation of the third harmonic. \textbf{Right:} Spectrogram (top), signaling the excitation of the third harmonic, and spectral broadening, and the reconstructed spatial profile of the different harmonic components (bottom), demonstrating clear focusing. }
  \label{fig:scalar_field}
\end{figure*}
We work perturbatively, assuming that the background solution is flat space. Thus, we take
\be\label{eq:perturbation_ekg}
    m=\varepsilon^2 m_2\, , \quad \beta=\varepsilon^2 \beta_2 \, , \quad \phi =\varepsilon \phi_1 + \varepsilon^3 \phi_3 \, ,
\ee
and analyze the dynamics of the leading nonlinear correction $\phi_3$ -- since the stress-energy tensor is quadratic in $\phi$, the second order correction to the scalar field is trivial. We follow the evolution of an ``imploding'' wave packet, consisting on the superposition of three driving frequencies $\Omega_i$, with $i=1,2,3$, of the form
\be
u_0(r)=\frac{1}{r}\sum_{i=1}^3\cos\Bigl(\Omega_i (r-r_0)\Bigr)e^{-\frac{1}{2}(\frac{r-r_0}{\sigma})^{2}} \,,\label{eq:Gaussian_initial_condition} 
\ee
and $u_1=u_0' + u_0/r$, so the wavepacket is moving towards the origin of coordinates. Note that a similar framework was used to understand BH formation in the full nonlinear regime~\cite{Choptuik:1992jv}. Here we focus instead on the nonlinear dynamics on the \emph{dispersive} regime, far enough away from the critical threshold of BH formation so that perturbation theory remains valid. We plan to examine nonlinear dynamics closer to the critical solution in a subsequent work (see also Refs.~\cite{Balasubramanian:2014cja,Yang:2015jja}) for work in this direction, in the context of confined-like geometries. We note that we have reproduced quantitatively all results that we present with a nonlinear code in the regime of small initial amplitude. Details of the numerical implementation, including convergence tests, are given in the Supplemental Material~\cite{supplemental}.
In the following, we demonstrate evidence for three foundational nonlinear effects with clear analogies in, e.g., nonlinear optics~\cite{Boyd_book}: (i) higher harmonic generation, (ii) focusing and spectral broadening, and (iii) modified peeling properties at higher frequencies.

An example of the time-evolution of our initial data is shown in the left panel of Fig.~\ref{fig:scalar_field}, for $\Omega_1=\Omega_2=\Omega_3\equiv \Omega_d=1$. The central panel shows the Fourier transform of the fields $\phi_1,\phi_3$ at $r=200$ at early times (when they are ingoing), and late times (outgoing). While the linear field is quasi-monochromatic, the nonlinear correction has a higher-harmonic component at the third harmonic $3\Omega_d$. The amplitude of this component changes between the ingoing and outgoing pieces, signaling that nonlinear harmonic generation is enhanced in the focusing region (near $r=0$), where the field density is largest. Finally, the right panel of Fig.~\ref{fig:scalar_field} shows the spectrogram of $\phi_3$, where spectral broadening can be seen in the spectral line at the driving frequency $\Omega_d$. In order to examine this further, we define the reconstructed radial profile of each frequency component as 
\begin{equation}
    \psi_{i,\omega} = \mathcal{F}_r^{-1}\Bigl[W_\omega \mathcal{F}_r(r\phi_i)\Bigr] \, , \qquad i=1,3 \, , \label{eq:psi_def}
\end{equation}
where $\mathcal{F}_r$ denotes the (radial) Fourier transform at some fixed instant in time, and $W_\omega$ is a window function that zeroes out frequency contributions far from the frequency $\omega$. The bottom-right panel of Fig.~\ref{fig:scalar_field} then shows the focusing of $\psi_{3,\Omega_d}$ and, more enhanced, of $\psi_{3,3\Omega_d}$, compared to the radial profile of $\psi_{1,\Omega_d}$ (which matches the initial data). Higher-harmonic generation is a direct consequence of mode coupling in nonlinear systems. Spectral broadening (spatial focusing) is a well-known phenomena in systems also featuring third harmonic generation such as nonlinear crystals --- the Kerr effect~\cite{Boyd_book} makes the refraction index amplitude-dependent, leading to focusing of waves near the regions with larger amplitude. Here we report for the first time the systematic observation of a gravitational Kerr effect (the effect itself had been observed in non-asymptotically flat geometries~\cite{Bizon:2011gg}).

\begin{figure}[t]
    \centering
    \includegraphics[width=\columnwidth]{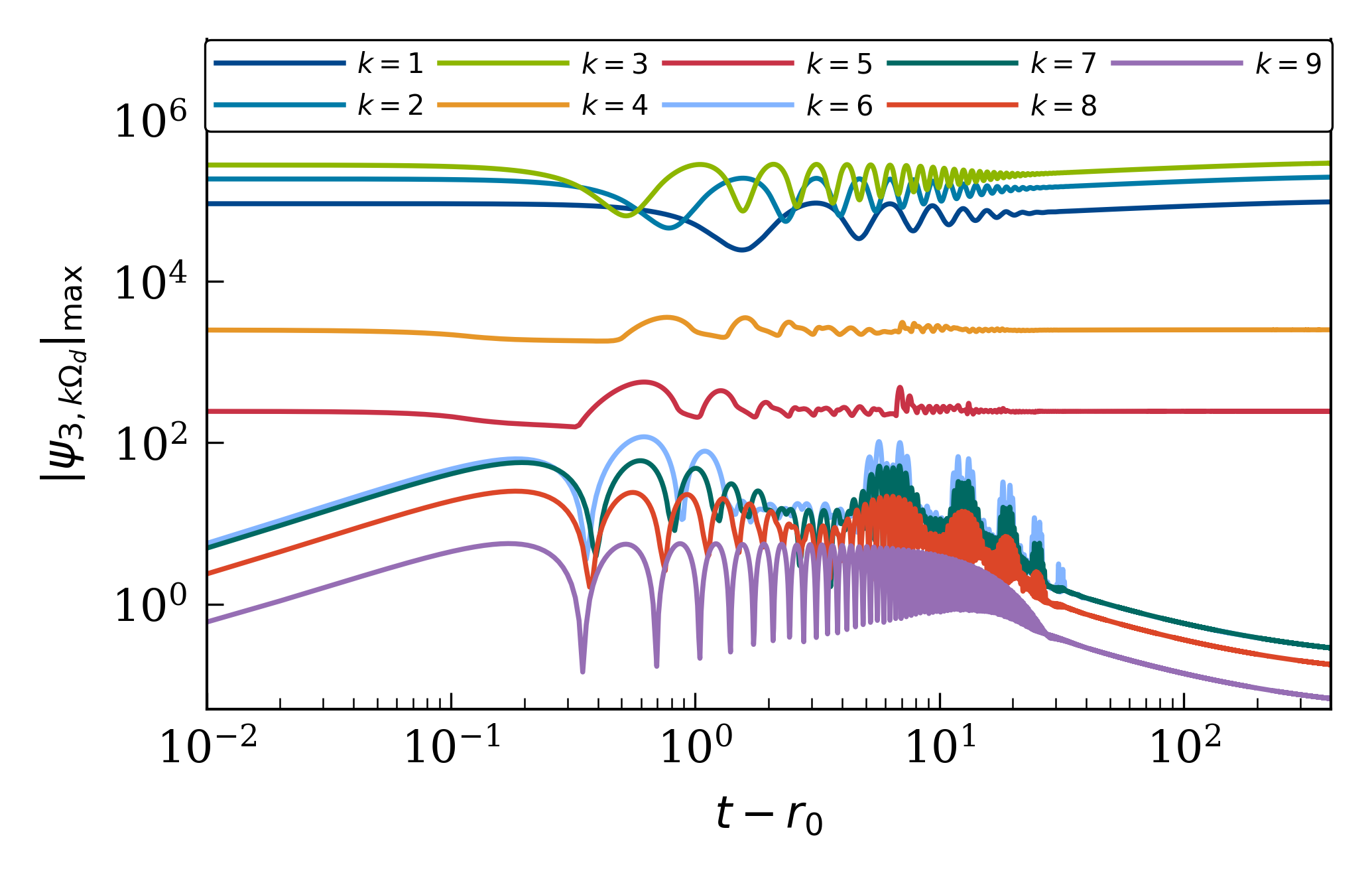}
    \caption{Decay properties of each of the higher harmonics generated nonlinearly, with frequencies $\omega_i$, for three-harmonic initial data with $(\Omega_1,\Omega_2,\Omega_3)=(1,2,3)$.  
    \label{fig:peeling}}
\end{figure}
Finally, we observe that higher harmonics generated in the scattering of pulses exhibit different decay properties than the expected $\phi \sim e^{i\omega r}/r$ behavior at large distances. Fig.~\ref{fig:peeling} shows, for initial data containing three different harmonics, that the different frequency components (extracted using~\eqref{eq:psi_def}) showcase three different behaviors: (i) the amplitude of the driven harmonics $\omega=\Omega_{1,2,3}$ have a slow (logarithmic) growth at late times,  (ii) the higher harmonics that receive contributions from multiple channels, e.g. $4=2\Omega_1+\Omega_2 = \Omega_3+\Omega_2-\Omega_1$, asymptote to a constant value, and (iii) the highest harmonics decay as $1/\log|t-r_0|$. This behavior is accurate to percent-level at late enough times.

A nontrivial spatial dependence of higher harmonics is a well-known phenomena, in particular in the context of fluid dynamics. For example, waves ``breaking'' in the ocean can be understood by studying the nonlinear equation for the fluid displacement $\xi$ on a medium with constant sound speed $c_0$ and adiabatic index $\gamma$~\cite{Beyer:1974,Gurbatov_book}.
\be
\ddot{\xi}-\frac{c_0^2}{(1+\xi')^{\gamma+1}}\xi''=0\,,
\ee
with dots standing for time derivatives and primes for spatial derivatives. This is a nonlinear wave equation, whose first nonlinear correction can be solved analytically. If the linearized solution is a traveling wave with frequency $\omega$, the nonlinear correction has components with frequency $2\omega$ which grow linearly with the distance $x$ -- perturbation theory fails at large distances, signaling the breaking of initially small waves. Although more work is needed to understand the ``peeling'' properties of higher harmonics in the setup we consider, we may be unveiling an inverse phenomena to ocean wave breaking: propagation of waves in GR suppresses the nonlinearly generated harmonics. 

\noindent \textbf{\textit{Scattering of GWs off a BH.}} Consider now the leading nonlinear correction to the scattering of GWs off a nonrotating BH. In particular, we consider a perturbative expansion where the spacetime metric is $g_{ab}=g^{\rm Sch}_{ab} + \varepsilon h^{(1)}_{ab} + \varepsilon^2 h^{(2)}_{ab}$, with $g^{\rm Sch}=-fdt^2+f^{-1}dr^2 + r^2d\Omega^2$ the Schwarzschild metric in areal radius coordinates, with $f=1-2M/r$. For simplicity, we take the linear perturbation $h^{(1)}_{ab}$ to be purely of axial (odd) parity. It is generated by a single master variable $\psi^{(1)}$, which is a solution to the Regge-Wheeler equation~\cite{Regge:1957td} with scattering boundary conditions
\begin{equation}
    \psi^{(1)} = \begin{cases}
        e^{i\Omega_d r_*} + \mathcal{O}(1-2M/r) \, , \quad &r \to 2M \, , \\
        A_{\rm in}e^{-i\Omega_d r_*} + A_{\rm out} e^{i\Omega_d r_*} + \mathcal{O}(r^{-1}) \, , \quad &r\to\infty \, ,
    \end{cases}
\end{equation}
where $A_{\rm in(out)}\equiv A_{\rm in(out)}(\Omega_d,\ell)$ are the amplitudes of incoming and outgoing waves, and we normalize the field such that the transmitted part onto the horizon is exactly unity. Above, $r_*=r+2M\log(r/2M-1)$ is the usual tortoise coordinate. 

The framework to study second-order perturbations in Schwarzschild is now largely developed~\cite{Gleiser:1995gx,Ioka:2007ak, Nakano:2007cj, Brizuela:2006ne, Brizuela:2009qd,Pazos:2010xf,Bucciotti:2024zyp, Bucciotti:2024jrv, Perrone:2023jzq, Bourg:2024jme, Bourg:2025lpd}. However, previous work focused mostly on the quadratic coupling of quasinormal modes, and the nonlinear scattering problem has remained open. In general, the second-order metric perturbations are also generated by a master scalar $\psi^{(2)}$. When the linear perturbation has a single well defined angular harmonic dependence, characterized say, by the angular number $\ell$, and is odd parity, then $h^{(2)}_{ab}$ is a purely even metric perturbation~\cite{Bucciotti:2024jrv}. Therefore, $\psi^{(2)}$ satisfies a sourced Zerilli equation
\begin{equation}\label{eq:2nd_order_z}
    \frac{d^2\psi^{(2)}}{dr_*^2}+(4\Omega_d^2-V_{2\ell}^{\rm e})\psi^{(2)} = S(\psi^{(1)},\psi^{(1)}) \, ,
\end{equation}
with $V_{2\ell}^{\rm e}$ the Zerilli potential 
\begin{equation}
        V_{2\ell}^{\rm e} = \frac{f}{\Lambda^2}\Bigl[\frac{\mu^4}{r^2}\Bigl(\ell(\ell+1)+\frac{6M}{r}\Bigr) + \frac{36M^2}{r^4}\Bigl(\mu^2 + \frac{2M}{r}\Bigr)\Bigr] \, , 
\end{equation}
where we introduced $\mu^2=(\ell+2)(\ell-1)$ and $\Lambda = \mu^2+6M/r$. The source term $S(\psi^{(1)},\psi^{(1)})$ does not decay sufficiently fast for $\psi^{(2)}$ to converge at large distances. Nevertheless, the metric perturbation is expected to be regular -- the divergence of $\psi^{(2)}$ is canceled by the divergences of the source. An alternative approach is to define a regularized master variable $\tilde{\psi}^{(2)} = \psi^{(2)}+\Upsilon_\ell$, with 
\begin{equation}
    \Upsilon_{\ell=2} = \Bigl(- \frac{\omega^2 r}{105\sqrt{\pi}} +\frac{M\omega^2}{315\sqrt{\pi}} + \frac{261-4M^2\omega^2}{3780\sqrt{\pi} \, r}\Bigr) (\psi_o^{(1)})^2 \, ,
\end{equation}
where we have specified $\ell=2$ for simplicity. The procedure used to obtained this regularization function can easily be generalized for any value of the angular momentum parameter (the $\ell=3,4$ cases are given in the Supplemental Material~\cite{supplemental}). With this regularization scheme, the new source term is
\begin{equation}
    \mathfrak{S}_\ell = S + \frac{d\Upsilon}{dr_*^2} + (4\Omega_d^2-V_{2\ell}^{\rm e})\Upsilon \xrightarrow{r\to\infty} r^{-2}  \, ,
\end{equation}
which decays sufficiently fast at large distances, and it is guaranteed to be regular at the horizon, provided that the original source term is. The asymptotic behavior of the (regularized) source term is shown in the SM. 

\begin{figure}
    \centering
    \includegraphics[width=\columnwidth]{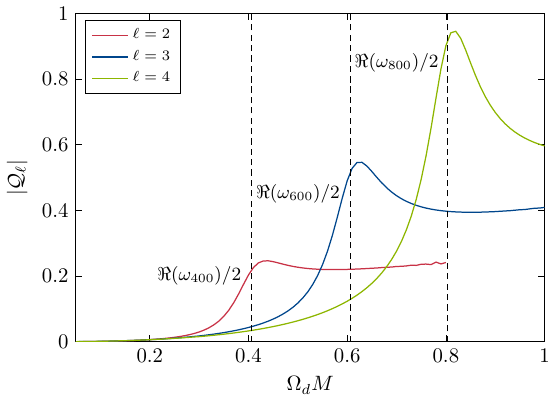}
    \caption{Susceptibility~\eqref{eq:q_def} of the excitation of the higher harmonic as a function of the dimensionless driving frequency $\Omega_d M$ of the axial quadrupolar with $\ell=2,3,4$ in red, blue, and green, respectively. At low frequencies the susceptibility vanishes, in agreement with the flat space limit~\eqref{eq:flat_space_q}. The susceptibility peaks close to the resonance condition $2\Omega_d=\Re\omega_{\ell 00}$, indicated with dashed lines. We have cut the $\ell=2$ line since the integral exhibits poor convergence at high frequencies.}
    \label{fig:qfactor}
\end{figure}

We solve the second-order regularized Zerilli equation~\eqref{eq:2nd_order_z} with the method of variation of parameters, enforcing quadratic perturbations $\tilde{\psi}^{(2)}$ to be ingoing at the horizon and outgoing at infinity. Let $\psi_{\rm in}, \psi_{\rm up}$ be homogeneous solutions of the Zerilli equation with the behaviour:
\begin{equation}
    \psi_{\rm in} \overset{r\to 2M}{\sim }  e^{-2i\Omega_d r_*} \, , \qquad \psi_{\rm up} \overset{r\to \infty}{\sim } e^{2i\Omega_d r_*} \, .
\end{equation}
The driven solution oscillating with frequency $2\Omega_d$ can be compactly written as~\cite{Cardoso:2025npr}
\begin{equation}
    \begin{aligned}
        \tilde{\psi}^{(2)} =& \frac{\psi_{\rm in}}{4i\Omega_d A_{\rm in}(2\Omega_d,2\ell)}\int_{r_*}^\infty \mathfrak{S}_\ell\psi_{\rm up} dr_* \\
        &+\frac{\psi_{\rm up}}{4i\Omega_d A_{\rm in}(2\Omega_d,2\ell)}\int_{-\infty}^{r_*} \mathfrak{S}_\ell \psi_{\rm in} dr_* \, .
    \end{aligned}
\end{equation}
At large distances, 
\begin{equation}
    \tilde{\psi}_e^{(2)} = \mathcal{Q}_\ell(\Omega_d) A_{\rm out}(\Omega_d,\ell)^2  e^{2i\Omega_d r_*} \, , \qquad r_* \to \infty \, , 
\end{equation}
where we have defined the nonlinear susceptibility as 
\begin{equation}\label{eq:q_def}
        \mathcal{Q}_\ell(\Omega_d) =\frac{\int_{-\infty}^\infty \mathfrak{S}_{\ell}[\psi^{(1)}_{\rm o}] \psi_{\rm in} dr_*}{4i\Omega_d A_{\rm in}(2\Omega_d,2\ell)A_{\rm out}(\Omega_d,\ell)^2} \,.
\end{equation}
The susceptibility $\mathcal{Q}_\ell$ decomposes into a ``geometric'' term (the denominator), akin to QNM excitation factors, and a source-dependent term (the numerator). We have computed the nonlinear susceptibility by numerically integrating~\eqref{eq:q_def} (convergence of results is demonstrated in SM). Our results are summarized in Fig.~\ref{fig:qfactor}. In particular, we have computed the nonlinear susceptibility for $\ell=2,3,4$, and we find that this is peaked close to the resonance condition $2\Omega_d = \Re \omega_{\ell 0 0}$, with $\omega_{\ell m 0}$ the frequency of the fundamental quasinormal mode of the Schwarzschild BH. The agreement with this resonance condition becomes better for larger values of $\ell$. At high frequencies $\Omega_d M\gg 1$, it seems that $|\mathcal{Q}_\ell|\to \mathrm{const}$.
WKB techniques might be useful in informing us about this regime, which is challenging to study numerically. Graviton scattering amplitude calculations will also be valuable at validating and providing analytical expressions for the low-frequency limit of our results~\cite{Bautista:2021wfy, Bautista:2022wjf, Bautista:2026qse, Bjerrum-Bohr:2025bqg, Bjerrum-Bohr:2026fhs, Ivanov:2024sds, Ivanov:2026icp}.

In the flat space limit $M\to0$, the source term is simply
\begin{equation}
\mathfrak{S}_{\ell=2} = \frac{18 }{35\sqrt{\pi}r^2}\psi_o^{(1)}\Bigl(\frac{3\psi_o^{(1)}}{r} -(\psi_o^{(1)})'\Bigr)  \, .
\end{equation}
The flat space source does not depend directly on the driving frequency and it decays as $r^{-3}$ at large distances. This signals that the nonlinear susceptibility vanishes exactly in the flat space limit: \emph{GWs do not couple quadratically in Minkowski}~\cite{Mendonca:2002fm, Servin:2003gv, Mendonca:2003qm, Ianniccari:2025nkf}. Indeed, regular homogeneous solutions at the origin are ($\ell=2$ for us, for the linear component) $\psi^{(1)} = A\sqrt{\Omega_d \,r}J_{\ell+1/2}(\Omega_d r)$, where the normalization has been fixed so that $A_{\rm in}=A_{\rm out}=A/\sqrt{2\pi}$. Putting all the pieces together, one finds
\begin{equation}
    \begin{aligned}
        \tilde{\psi}_e^{(2)} =& \frac{9A^2}{1120\pi^{3/2}r^5\Omega_d^4}\Biggl[48(6+\Omega_d^2r^2) \\
        &-2\Bigl(144-159\Omega_d^2r^2+8\Omega_d^4r^4\Bigr)\cos(2\Omega_d r) \\
        &+3\Omega_d r\Bigl(36\Omega_d^2r^2-157\Bigr)\sin(2\Omega_d r)\Biggr] \, .\label{eq:flat_space_q}
    \end{aligned}
\end{equation}

Remarkably, the quadratic driven GWs decay as $\tilde{\psi}^{(2)}= \mathcal{O}(r^{-1})$, while $\psi^{(1)}= \mathcal{O}(1)$ at large distances, confirming that the nonlinear susceptibility vanishes $\mathcal{Q}_{\ell=2}=0$. This is also confirmed in the $\Omega_d M\to 0$ limit of Fig.~\ref{fig:qfactor}. Nevertheless, there is a higher harmonic which is excited in~\eqref{eq:flat_space_q} -- it simply does not propagate all the way to null infinity. This seems in agreement with our observations in the scattering of spherical scalar waves in flat space, but warrants further research. 

We note that time evolutions of the quadratic equations -- not reported here -- show the same features as before: higher harmonics generation, focusing and spectral broadening. With this is mind we turn to a full nonlinear problem.

\noindent \textbf{\textit{Nonlinearities in BH mergers.}} Motivated by the previous findings, we have also examined the nonlinear gravitational dynamics during BH mergers. For this purpose, we have evolved the approximately quasi-circular, nonspinning, equal-mass
BH binary configuration labeled R1 in Table I of Ref.~\cite{Baker:2006yw} using the \textsc{lean} code
\cite{Sperhake:2006cy}. \textsc{lean} is based on the
\textsc{cactus} computational toolkit \cite{Allen:1999},
employs \textsc{carpet} \cite{Schnetter:2003rb} for
mesh refinement and computes apparent horizons with
\textsc{ahfinderdirect} \cite{Thornburg:1995cp,Thornburg:2003sf}. This simulation
evolves the Einstein equations with the Baumgarte-Shapiro-Shibata-Nakamura (BSSN) formulation \cite{Shibata:1995we,Baumgarte:1998te}
and the moving-puncture method \cite{Baker:2005vv,Campanelli:2005dd}
on a computational domain of size $512\,M$ with 7 refinement levels
and resolution $M/48$ on the innermost grid, where $M$ denotes
the total BH mass. This corresponds to the highest-resolution
employed for this configuration in Sec.~III of Ref.~\cite{Sperhake:2006cy}
which estimates errors in the GW energy and amplitude at the percent
level.

\begin{figure}[t]
    \centering
        \includegraphics[width=\columnwidth]{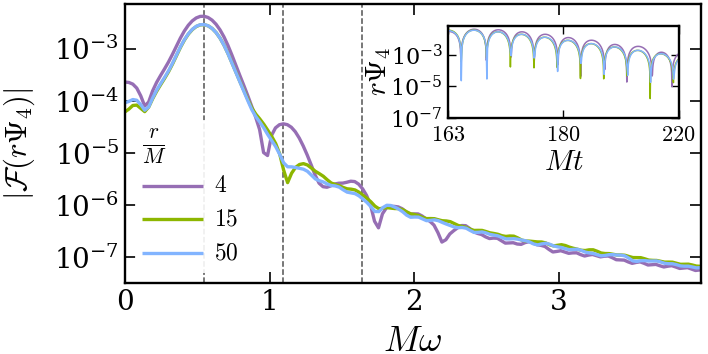}
    \caption{Fourier transform of the quadrupole $r\Psi_4$ extracted at three different observing radii $r_{\rm obs}/M=4,15,20$. The dashed lines correspond to $k\Re[\omega_{220}]$ for $k=1,2,3$, with $\omega_{220}$ the fundamental mode of the remnant BH. The inset shows the time-domain response, where the waveforms are aligned to the peak of $r_{\rm obs}=4M$. The Fourier transform is only applied in the time window shown in the inset, corresponding to the merger-ringdown phase. 
    }
    \label{fig:bbh}
\end{figure}
We extract the leading spherical harmonic multipole $(\ell,m)=(2,2)$ of the curvature scalar $\Psi_4$ at worldtubes of fixed radii close to the merger region. Then we take the Fourier transform of the GWs emitted during the merger-ringdown stage, as shown in Fig.~\ref{fig:bbh}. The signal is peaked at the fundamental mode frequency of the remnant BH, $M\Re\omega\simeq 0.57$, for all extraction radii. However, the purple line, corresponding to extracting closest to the strong-field region, has a clear peak at the second harmonic $\omega\sim 2\Re\omega_{220}$, and a weaker peak at the third harmonic $\omega\sim 3\Re\omega_{220}$. These peaks -- which should not be mistaken with quadratic QNMs, which are indeed present at future null infinity -- are absent when extracting further away from the BH. This suggests that nonlinear higher harmonic excitation is present during BH mergers, but it may have a nontrivial spatial dependence -- in particular, nonlinearities may be ``washed away'' as they propagate in spacetime, leading to an approximately linear signal at future null infinity, in good agreement with perturbation theory results.

\noindent \textbf{\textit{Discussion.}} Nonlinear phenomena are ubiquitous in nature. Recently, these have been brought to the spotlight, motivated by the unexpected success of perturbation theory in describing the GW emission during BH mergers~\cite{Gleiser:1995gx, Gleiser:1996yc,Buonanno:2006ui,Sperhake:2011ik,Nagar:2013sga,Green:2013zba,Yang:2014tla,vandeMeent:2020xgc,Ma:2025rnv}, and also by unveiling detectable nonlinear harmonics in the relaxation of BHs to equilibrium~\cite{Zlochower:2003yh,London:2014cma,Cheung:2022rbm, Mitman:2022qdl,Yi:2024elj,Lagos:2024ekd}. However these effects are part of a bigger picture, whose systematic study we first tackle in this work. Our message is clear: gravitational systems are prone to nonlinear effects, namely higher harmonic generation, 
but also spatial focusing and spectral broadening -- a gravitational analogue to the Kerr effect.  

These findings are important to understand the gravitational interaction at large, and also when discussing mergers. The peeling properties of some nonlinear terms suggest that the near horizon region could be ``boiling'' with higher harmonic content, whereas the GW at our detector shows only a smooth, low-harmonic content. This would then challenge us to study mergers also in the optical channel, for example. Likewise, the anatomy of binary mergers~\cite{Buonanno:2006ui,Berti:2007fi} may need or benefit from improvements concerning both the delays or focusing of nonlinear signals: in particular, an interpretation based solely on nonlinear ringdown modes starting from the merger may simply not reflect the intricate nonlinear phenomena at play, as we showed. Identification of these effects, however, requires that we know in advance how the signal would look like in absence of nonlinearities. Extreme mass ratio systems seem to be a good candidate, where one has a controlled way (mass ratio) to go smoothly from linearized to full nonlinear theory.

\begin{acknowledgments}
%
We thank Gregorio Carullo and Luis Lehner for conversations that started this project. We are indebted to Bruno Bucciotti and Adrien Kuntz for numerous discussions and comments and for pointing out an error in the original version of source term to second order vacuum fluctuations. 
The Center of Gravity is a Center of Excellence funded by the Danish National Research Foundation under grant No. DNRF184.
We acknowledge support by VILLUM Foundation (grant no. VIL37766).
V.C.\ is a Villum Investigator.  
V.C. acknowledges financial support provided under the European Union’s H2020 ERC Advanced Grant “Black holes: gravitational engines of discovery” grant agreement no. Gravitas–101052587. 
Views and opinions expressed are however those of the author only and do not necessarily reflect those of the European Union or the European Research Council. Neither the European Union nor the granting authority can be held responsible for them.
This project has received funding from the European Union's Horizon 2020 research and innovation programme under the Marie Sklodowska-Curie grant agreement No 101007855 and No 101131233.
This work is supported by Simons Foundation International \cite{sfi} and the Simons Foundation \cite{sf} through Simons Foundation grant SFI-MPS-BH-00012593-11.
The Tycho supercomputer hosted at the SCIENCE HPC center at the University of Copenhagen was used for supporting this work.
This work has been supported by STFC Research Grant No. ST/V005669/1. We acknowledge support by the NSF Grant No.~PHY-090003, DiRAC projects ACTP284 and ACTP238, STFC capital Grants No.~ST/P002307/1, No.~ST/R002452/1, No.~ST/I006285/1 and No.~ST/V005618/1, STFC operations Grant No.~ST/R00689X/1. Computations were partially done on the CSD3 and Swirles (Cambridge), Cosma (Durham), Stampede2 (TACC) and Expanse (SDSC) clusters.

\end{acknowledgments}

\clearpage
\bibliography{biblio}

\clearpage
\begin{appendix}
\onecolumngrid
\section*{Supplemental Material}

\section{Einstein-Klein-Gordon}

\begin{figure}[H]
    \centering
    \includegraphics[width=0.5\linewidth]{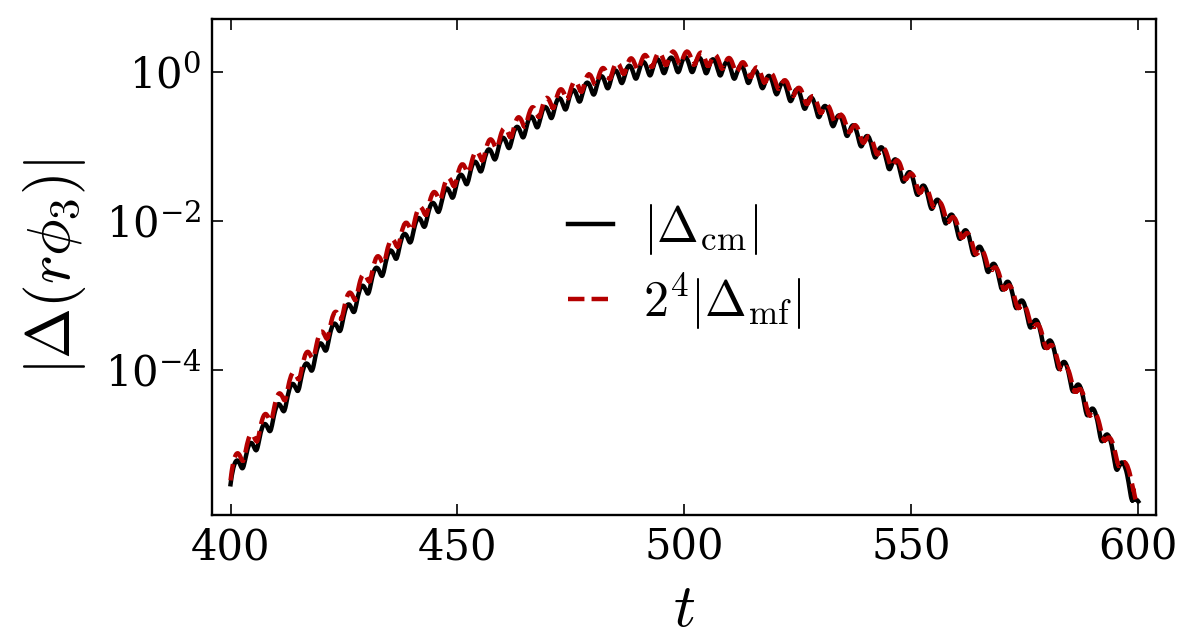}
    \caption{Residuals in the envelope of the nonlinear correction to the scalar field $r\phi_3$ observed at $r_{\rm obs}=300\Omega_1$,  between coarse ($\Delta r=0.1$) and medium ($\Delta r=0.05$) resolutions, in black, and between medium and fine ($\Delta r = 0.025$) resolutions, rescaled by the expected fourth order convergence factor, in red. The two lines overlap, signaling fourth order convergence of our results.
    \label{fig:convergence_test}
    }
\end{figure}

In the first part of this letter, we study perturbatively the Einstein-Klein-Gordon system in spherical symmetry for a massless scalar field. Using the notation of~\eqref{eq:perturbation_ekg}, the Einstein equations become
\begin{equation}
    \begin{aligned}
        m_2^{'}&=\kappa r^{d-1}(\dot{\phi}^2_1+{\phi^{'}}^{2}_1)\\
        \beta ^ {'}_2&=\frac{(d-2)m_2}{r^{d-1}}\\
        \square \phi_1 =& 0 \, , \qquad \square\phi_3=2\beta_2\ddot{\phi}_1+\dot{\beta}_2\dot{\phi}_1+\beta ^{'}_2\phi^{'}_1 \, , 
    \end{aligned}
\end{equation}
together with the constraint $\dot{m}_2=2\kappa r^{d-1}\dot{\phi}_1\phi^{'}_1$, where primes denote radial derivatives, and dots denote time derivatives. We evolve these equations writing the wave equations in first-order form for initial data of the form~\eqref{eq:Gaussian_initial_condition} using the method of lines. Spatial derivatives are approximated with 4th order finite differences, and we evolve forwards in time with an explicit fourth order Runge-Kutta algorithm. Our code is implemented in \texttt{Python}, and is available via~\cite{web:CoG}. The simulations presented in the same text are performed with spatial resolution $\Delta r =0.014 $. The convergence of our results is shown in Fig.~\ref{fig:convergence_test}. We also note that we use a Hann window to smoothen the data before taking Fourier transforms, and that the window function used in~\eqref{eq:psi_def} is a rectangular window in momentum space -- we have tested that using different window functions does not alter our results. 

\section{Regularized Source for $\ell=2,3,4$}

Here we detail the regularized source for the multipoles considered in Fig.~\ref{fig:qfactor}. In all cases we consider the angular channel: $(\ell,0)\times(\ell,0) \to (2\ell,0)$, for $\ell=2,3,4$, where the linear perturbations have purely axial parity (and hence the higher harmonics have purely even parity). The original source scales as $\mathcal{O}(r-2M)$ at the horizon, but it grows rapidly at large distances. We regularize this behavior by redefining the master variable as $\tilde{\psi}^{(2)}=\psi^{(2)}+\Upsilon_\ell$. The function $\Upsilon_\ell$ always takes the same form:
\begin{equation}
    \Upsilon_\ell = \Bigl(a_\ell r + b_\ell + \frac{c_\ell}{r}\Bigr)(\psi_o^{(1)})^2 \, ,
\end{equation}
and the coefficients are given by 
\begin{equation}
    \begin{aligned}
        \ell = 2: \qquad & a_2 = -\frac{\omega^2}{105\sqrt{\pi}}, \quad b_2 = \frac{M\omega^2}{315\sqrt{\pi}}, \quad c_2 = \frac{261-4M^2\omega^2}{3780\sqrt{\pi}}  \, , \\
        \ell = 3: \qquad & a_3 = -\frac{5\omega^2}{308\sqrt{13\pi}}, \quad b_3 = \frac{3M\omega^2}{1232\sqrt{13\pi}} , \quad c_3 =\frac{(22000-9M^2\omega^2)\sqrt{13}}{320320\sqrt{\pi}}  \, , \\
        \ell = 4: \qquad & a_4 = -\frac{14\omega^2}{1287\sqrt{17}} , \quad b_4 = \frac{2M\omega^2}{2145\sqrt{17\pi}}, \quad c_4 = \frac{(109025-4M^2\omega^2)\sqrt{17}}{850850\sqrt{\pi}} \, .
    \end{aligned}
\end{equation}
For the sake of reproducibility, we write here the regularized source for $\ell=2$, which is simply 
\begin{equation}
    \begin{aligned}
        \mathfrak{S}_{\ell=2} =& -\frac{M (r-2 M)^2 \left(4 M^3 \omega ^2+12 M^2 r \omega ^2+117 M+162 r\right) }{1890 \sqrt{\pi } r^3 (M+3 r)^2} [(\psi^{(1)}_o)']^2\\
        &-\frac{(r-2 M)^2 \left(-4 M^4 \omega ^2-24 M^3 r \omega ^2+639 M^2+3186 M r+4374 r^2\right) }{945 \sqrt{\pi } r^4 (M+3 r)^2}\psi^{(1)}_o(\psi^{(1)}_o)'\\
        &+\Biggl[\frac{-80 M^6 \omega ^2-240 M^5 r \omega ^2-4 M^4 \left(r^4 \omega ^4-29 r^2 \omega ^2-927\right)-6 M^3 r \left(2 r^4 \omega ^4-2 r^2 \omega ^2-3393\right)}{1890 \sqrt{\pi } r^5 (M+3 r)^2}\\
        &\qquad-\frac{9 M^2 r^2 \left(r^2 \omega ^2+322\right)-216 M r^3 \left(r^2 \omega ^2+262\right)+26244 r^4}{1890 \sqrt{\pi } r^5 (M+3 r)^2}\Biggr](\psi^{(1)}_o)^2\, .
    \end{aligned}
\end{equation}
The derivation of this source, as well as the regularized sources also for $\ell=3,4$, can be found in~\cite{web:CoG}.

\section{Nonlinear Susceptibility in Schwarzschild}

\begin{figure*}[h]
\includegraphics[width=0.48\linewidth]{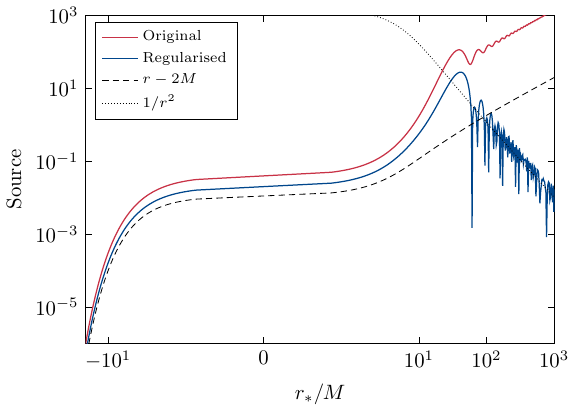}
\includegraphics[width=0.48\linewidth]{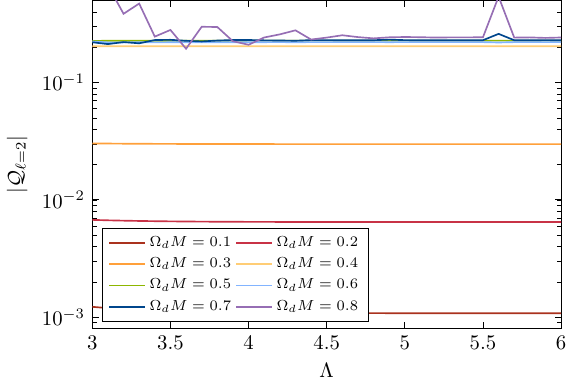}
  \caption{\textbf{Left:} Original source $S_{\rm ooe}$ (red) and regularized source $\mathfrak{S}_{\rm ooe}$ for $\ell=2$ and $\Omega_d M=0.1$, as a function of the tortoise coordinate $r_*$. The dashed and dotted lines represent the expected (regular) behavior of the source close to the horizon ($r_*\to-\infty$), and towards infinity ($r_*\to\infty$). \textbf{Right:} Convergence of the nonlinear susceptibility $\mathcal{Q}_{\ell=2}$ when increasing the regulator of the integral $\Lambda$, for different values of the driving frequency $\Omega_d$ as indicated in the legend. }
  \label{fig:details_q}
\end{figure*}

In order to compute the nonlinear susceptibility of higher harmonics in a Schwarzschild background, as defined in~\eqref{eq:q_def}, we need to accurately compute the integral of the numerator of~\eqref{eq:q_def}. In order to do so, we first integrate the homogeneous Regge-Wheeler and Zerilli equations with appropriate boundary conditions at finite radii
\begin{equation}
    \psi_{\rm in} = e^{-i\omega r_*}\sum_{n=0}^{N} a_n (r-2M)^n \, , \qquad  \psi_{\rm up} = e^{i\omega r_*}\sum_{n=0}^\infty \frac{b_n}{r^n} \, , 
\end{equation}
where the coefficients $a_n,b_n$ can be found analytically as functions of $\omega,\ell$, and we use typically $N=10$ to ensure that enough accuracy is achieved near the boundaries. The Regge-Wheeler and Zerilli equations are solved using a highly accurate $9$th order Verner method using \texttt{DifferentialEquations.jl}. The source term $\mathfrak{S}_{\rm ooe}[\psi_o^{(1)}]$ can be then evaluated, and the left panel of Fig.~\ref{fig:details_q} shows that it achieves the expected asymptotic behavior near the horizon, and at large distances. 

Despite the good asymptotic properties of the regularized source $\mathfrak{S}$, the integral in~\eqref{eq:q_def} is highly oscillatory and therefore numerically delicate. We use a regulator to study the convergence of the integral when increasing the size of our computational domain
\begin{equation}
    \mathcal{I}_\Lambda = \int_{2M+10^{-{\Lambda-1}}}^{10^{\Lambda-1}/\Omega_d} \mathfrak{S}_{\rm ooe}[\psi^{(1)}_{\rm o}] \psi_{\rm in} fdr \, .
\end{equation}
The integral $\mathcal{I}_\Lambda$ is performed using adaptive Gauss-Kronrod quadrature as implemented in the \texttt{Integrals.jl} package. Our code is fully implemented in \texttt{Julia} and available in~\cite{web:CoG}. The right panel of Fig.~\ref{fig:details_q} shows that $\mathcal{I}_\Lambda$ approaches a constant value when increasing $\Lambda$ for the frequencies considered in this work. Notice that the convergence with $\Lambda$ is worse for larger driving frequencies -- see also the oscillations towards the right edge of Fig.~\ref{fig:qfactor}. Other computational methods, possibly based on compact domains and horizon penetrating coordinates, may be better suited to perform this calculation accurately at high frequencies. 

\end{appendix}
\end{document}